\documentclass{PoS}
\usepackage{braket}
\usepackage{subfig}
\usepackage{float}
\usepackage{url}
\usepackage{amsmath}
\usepackage{amsfonts}
\usepackage{verbatim}
\usepackage{listings}
\usepackage{algorithm}
\usepackage{algorithmic}

\newcommand{\Tr}{\mbox{\rm Tr}}
\newcommand{\Real}{\mbox{\rm Re}}
\newcommand\T{\rule{0pt}{2.6ex}}
\newcommand\B{\rule[-1.2ex]{0pt}{0pt}}

\graphicspath{{figures/}}

\title{GPU implementation of a Landau gauge fixing algorithm}
\ShortTitle{GPU implementation of a Landau gauge fixing algorithm}

\author{Nuno Cardoso \\
CFTP, Departamento de F\'{\i}sica, Instituto Superior T\'ecnico,
Universidade T\'ecnica de Lisboa, Avenida Rovisco Pais 1, 1049-001 Lisbon, Portugal\\
E-mail: \email{nuno.cardoso@ist.utl.pt}}
\author{\speaker{Paulo J. Silva} \\
CFC, Departamento de F\'{\i}sica, Faculdade de Ci\^encias e Tecnologia, Universidade de Coimbra, 3004-516 Coimbra, Portugal\\
E-mail: \email{psilva@teor.fis.uc.pt}}
\author{Orlando Oliveira\\
CFC, Departamento de F\'{\i}sica, Faculdade de Ci\^encias e Tecnologia, Universidade de Coimbra, 3004-516 Coimbra, Portugal\\
E-mail: \email{orlando@teor.fis.uc.pt}}
\author{Pedro Bicudo\\
CFTP, Departamento de F\'{\i}sica, Instituto Superior T\'ecnico,
Universidade T\'ecnica de Lisboa, Avenida Rovisco Pais 1, 1049-001 Lisbon, Portugal\\
E-mail: \email{bicudo@ist.utl.pt}}

\abstract{
We discuss how the steepest descent method with Fourier acceleration for Laudau gauge fixing in lattice SU(3) simulations can be implemented using CUDA. 

The scaling of the gauge fixing code was investigated using a Tesla C2070 Fermi architecture, and compared with a parallel CPU gauge fixing code.
}

\FullConference{The 30th International Symposium on Lattice Field Theory\\
June 24 - 29, 2012\\
Cairns, Australia}

\begin{document}

\section{Introduction and motivation}

On the lattice, Landau gauge fixing is performed by maximising the functional
\begin{equation}
	F_U[g]=\frac{1}{4 N_cV}\sum_x\sum_\mu\Real\left[ \Tr\left(  g(x)U_\mu(x)g^\dagger(x+\mu) \right) \right] \, ,
\end{equation}
where $N_c$ is the dimension of the gauge group and $V$ the lattice volume, on each gauge orbit.
One can prove that picking a maximum of $F_U[g]$ on a gauge orbit is equivalent to demand the usual continuum Landau gauge condition $\partial_\mu A^a_\mu = 0$ and to require the positiveness of the Faddeev-Popov determinant. 
In the literature this gauge is known as the minimal Landau gauge.
The functional $F_U[g]$ can be maximised using the steepest descent method \cite{Davies:1987vs,Oliveira:2003wa}. 
However, when the method is applied to large lattices, it  faces the problem of critical slowing down, which can be attenuated by Fourier acceleration.

In the Fourier accelerated method, at each iteration one chooses
\begin{equation}
	g(x) = \exp\left[ \hat{F}^{-1} \frac{\alpha}{2}\frac{p^2_\text{max}a^2}{p^2a^2} \hat{F} \left( \sum_\nu \Delta_{-\nu}\left[U_\nu(x)-U_\nu^\dagger(x)\right]   -\text{trace} \right)\right]
\end{equation}
with
\begin{equation}
	\Delta_{-\nu}\left(U_\mu(x)\right) = U_\mu(x-a\hat{\nu})-U_\mu(x)\, ,
\end{equation}
$p^2$ are the eigenvalues of $\left(-\partial^2\right)$, $a$ is the lattice spacing and $\hat{F}$ represents a fast Fourier transform (FFT).
For the parameter $\alpha$, we use the recommended  value 
0.08 \cite{Davies:1987vs}.
In what concerns the computation of $g(x)$, for numerical purposes it is enough to expand the exponential to first order in $\alpha$, followed by a reunitarization, i.e., a projection to the SU(3) group.
The evolution and convergence of the gauge fixing process can be monitored by
\begin{equation}
	\theta = \frac{1}{N_c V}\sum_x \Tr\left[\Delta(x)\Delta^\dagger(x)\right]
\end{equation}
where
\begin{equation}
	\Delta(x) = \sum_\nu\left[ U_\nu(x-a\hat{\nu})-U_\nu(x) - \text{h.c.} - \text{trace} \right]
\end{equation}
is the lattice version of $\partial_\mu A_\mu=0$ and $\theta$ is the mean value of
$\partial_\mu A_\mu$ evaluated over all space-time lattice points per color degree of freedom.
In all the results shown below, gauge fixing was stopped only when $\theta \le 10^{-15}$.

\begin{algorithm}[!htb]
\begin{center}
\begin{algorithmic}[1] 
\STATE calculate $\Delta(x)$, $F_g[U]$ and $\theta$
\WHILE{$\theta \geq \epsilon$}
\FORALL{ element of $\Delta(x)$ matrix }
%\STATE apply FFT
\STATE apply FFT forward
\STATE apply $p^2_\text{max}/p^2$
%\STATE apply IFFT
\STATE apply FFT backward
\STATE normalize 
\ENDFOR
\FORALL{$x$}
\STATE obtain $g(x)$ and reunitarize
\ENDFOR
\FORALL{$x$}
\FORALL{$\mu$}
\STATE $U_\mu(x) \rightarrow g(x)U_\mu(x)g^\dagger(x+\hat{\mu})$
\ENDFOR
\ENDFOR
\STATE calculate $\Delta(x)$, $F_g[U]$ and $\theta$
\ENDWHILE
\end{algorithmic}
\end{center}
\caption{Landau gauge fixing using FFTs.}
\label{alg:fft}
\end{algorithm}

\section{Parallel implementation of the gauge fixing algorithm }

\subsection{CPU implementation}

The MPI parallel version of the algorithm was implemented in C++, using the machinery provided by the Chroma library \cite{Edwards2005}.
The Chroma library is built on top of QDP++, a library which provides a data-parallel programming environment suitable for Lattice QCD.
The use of QDP++ allows the same piece of code to run both in serial and parallel modes.
For the Fourier transforms, the code uses PFFT, a parallel FFT library written by Michael Pippig \cite{Pippig2011}.
Note that, in order to optimize the interface with the PFFT routines, we have compiled QDP++ and Chroma using a lexicographic layout.

\subsection{GPU implementation}

For the parallel implementation of the SU(3) Landau gauge fixing on GPU's \cite{Cardoso:2012pv}, we used version 4.1 of CUDA. 
For the 4D lattice, we address one thread per lattice site.
Although CUDA supports up to 3D thread blocks \cite{NVIDIA:cudac}, the same does not happen for the grid, which can be up to 2D or 3D depending on the architecture and CUDA version.
For grids up to 3D, this support happens only for CUDA version 4.x and for a CUDA device compute capability bigger than 1.3, i.e. at this moment only for the Fermi and Kepler architectures. Nevertheless, the code is implemented with 3D thread blocks and for the grid we adapted the code for each situation.
Since our problem needs four indexes, using 3D thread blocks (one for t, one for z and one for both x and y), we only need to reconstruct the other lattice index inside the kernel.

We use the GPU constant memory to put most of the constants needed by the GPU, like the number of points in the lattice, using \lstinline!cudaMemcpyToSymbol()!.
To store the lattice array in global memory, we use a SOA type array as described in \cite{Cardoso:2011xu}.
The main reason to do this is due to the FFT implementation algorithm. 
The FFT is applied for all elements of $\Delta(x)$ matrix separately. Using the SOA type array, the FFT can be applied directly to the elements without the necessity of copying data or data reordering.

On GPU's, FFT are performed using the CUFFT library by NVIDIA. Since there is no support for 4D FFTs, one has to combine four 1D FFTs, two 2D FFTs or 3D+1D FFTs. The best choice for our 4D problem is two 2D FFTs. 

In order to reduce memory traffic we can use the unitarity of SU(3) matrices and store only the first two rows (twelve real numbers) and reconstruct the third row on the fly when needed, instead of storing it.

\begin{table}[b]
\begin{center}
\begin{tabular}{|c|c|c|c|c|}
\hline
\T\B \textbf{kernel}	&	\multicolumn{2}{c|}{18real}	&	\multicolumn{2}{c|}{12real}	\\ \hline
\T\B per thread	&	load/store	& flop	&	load/store	& flop	\\ \hline
\hline
\T\B k1	&	0/1	&	20	&	0/1	&	20	\\ \hline
\T\B k2	&	144/20	&	505	&	96/14	&	841	\\ \hline
\T\B k3	&	2/2	&	0	&	2/2	&	0	\\ \hline
\T\B k4	&	3/1	&	2	&	3/1	&	2	\\ \hline
\T\B k5	&	18/18	&	153	&	12/12	&	153	\\ \hline
\T\B k6	&	162/72	&	1584	&	108/48	&	1962	\\ \hline
%\hline
\end{tabular}
\end{center}
\caption{Kernel memory loads/stores and number of floating-point operations (flop) per thread by kernel. The total number of threads is equal to the lattice volume. For kernel details see the text.}
\label{tab:memflop}
\end{table}

To perform Landau gauge fixing in GPUs using CUDA, we developed the following kernels:

\begin{itemize}
	\item k1: kernel to obtain an array with $p^2_\text{max}/p^2$.
	\item k2: kernel to calculate $\Delta(x)$, $F_g[U]$ and $\theta$. The sum of $F_g[U]$ and $\theta$ over all the lattice sites are done with the parallel reduction code in the NVIDIA GPU Computing SDK package, which is already an optimized code.
	\item k3: kernel to perform a data ordering.
	\item k4: apply $p^2_\text{max}/p^2$ and normalize.
	\item k5: obtain $g(x)$ and reunitarize.
	\item k6: perform  $U_\mu(x) \rightarrow g(x)U_\mu(x)g^\dagger(x+\hat{\mu})$.
\end{itemize}

In Table \ref{tab:memflop}, we show the number of floating-point operations and the number of memory loads/stores per thread by kernel. The number of floating-point operations using 2D plus 2D FFTs is given by $nx\times	ny\times nz\times nt\times 5( \log_2(nx\times ny)+\log_2(nz\times nt))$.

\begin{table}[!htb]
\begin{centering}
%\begin{tabular}{|c|c|c|c|}
\begin{tabular}{rl|rl}
\hline
\multicolumn{4}{c}{\T\B \textbf{NVIDIA Tesla C2070}}\tabularnewline
\hline
%\hline
\T\B Number of GPUs  & 1 & Global memory & 6144 MB\tabularnewline
\hline
\T\B CUDA Capability & 2.0 & Memory bandwidth (ECC off) & 148 GBytes/s\tabularnewline
\hline
\T\B Multiprocessors (MP) & 14 & Shared memory (per SM) & 48KB or 16KB\tabularnewline
\hline
\T\B Cores per MP & 32 & L1 cache (per SM) & 16KB or 48KB\tabularnewline
\hline
\T\B Total number of cores & 448 & L2 cache (chip wide) & 768KB\tabularnewline
\hline
\T\B  Clock rate & 1.15 GHz & Device with ECC support & yes \tabularnewline
\hline
\end{tabular}
\par\end{centering}
\caption{NVIDIA's graphics card specifications used in this work.}
\label{fig:nvidia_gpu_specs}
\end{table}

\section{Results}

In this section we show the performance of the GPU code, and compare with the parallel CPU code. 
The GPU performance results were performed on a NVIDIA Tesla C2070, Table \ref{fig:nvidia_gpu_specs}, and using version 4.1 of CUDA. The parallel CPU performance were done in the Centaurus cluster, at Coimbra. This cluster has 8 cores per node, each node has 24 GB of RAM, with 2 intel Xeon E5620@2.4 GHz (quad core) and it is equipped with a DDR Infiniband network.

\begin{comment}
\begin{figure}[!htb]
\begin{centering}
    \subfloat[\label{fig:time_32_3_ecc_off}]{
\begin{centering}
    \includegraphics[width=12cm]{time_32_3}
\par\end{centering}}

    \subfloat[\label{fig:time_32_3_ecc_on}]{
\begin{centering}
    \includegraphics[width=12cm]{time_32_3_ecc_on}
\par\end{centering}}
\par\end{centering}
    \caption{Time (s) per 1000 iterations, \cite{Cardoso:2012pv}. \protect\subref{fig:time_32_3_ecc_off} with ECC Off and \protect\subref{fig:time_32_3_ecc_on} with ECC On. sp: single precision, dp: double precision, 18real: full SU(3) matrix, 12real: 12 real parametrization, tex: using texture memory and cache: using L1 and L2 cache memory.}
    \label{fig:time_32_3}
\end{figure}
\end{comment}
\begin{figure}[!htb]
\begin{centering}
    \subfloat[\label{fig:perf_32_3_ecc_off}]{
\begin{centering}
    \includegraphics[width=12cm]{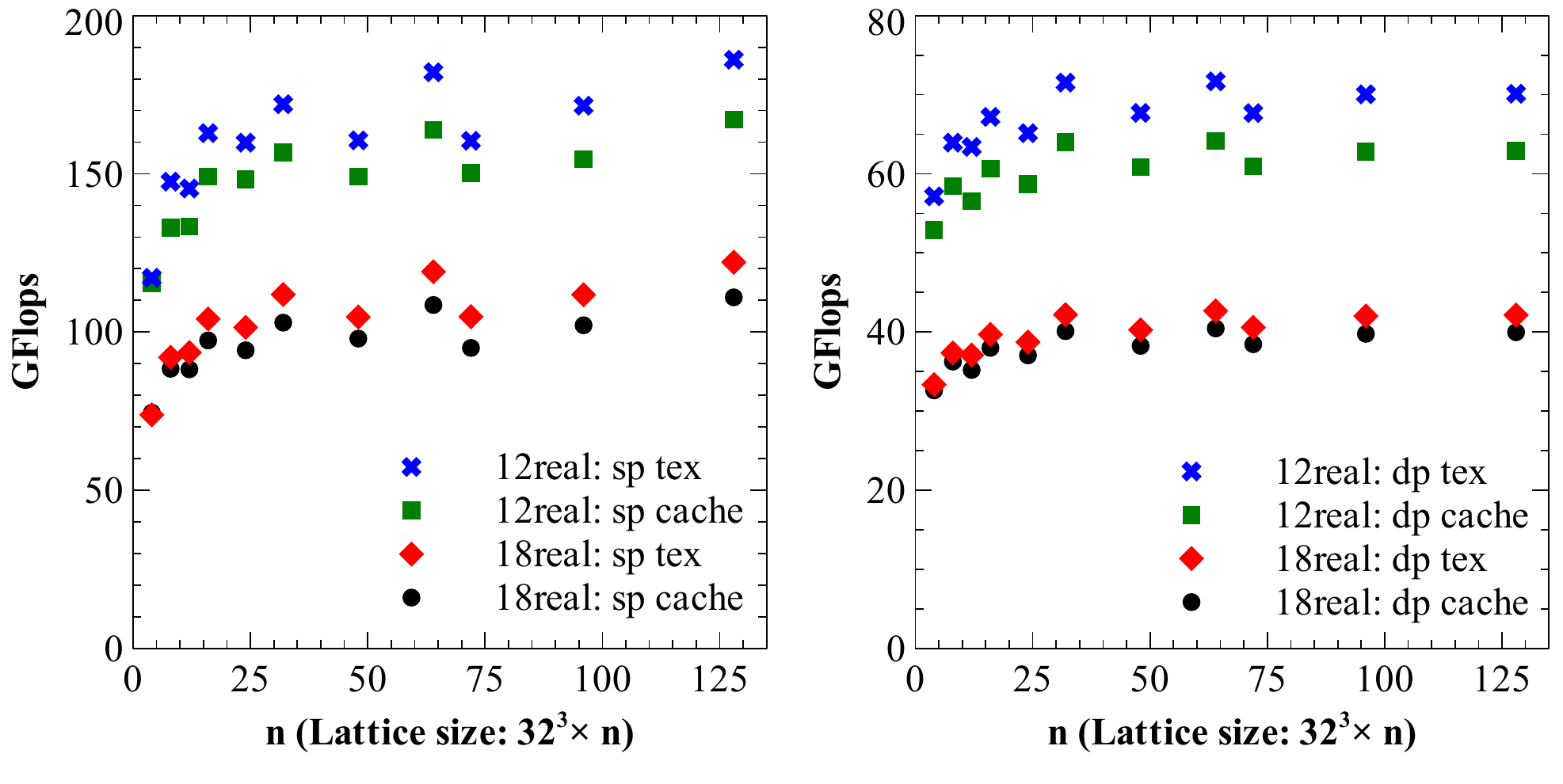}
\par\end{centering}}

    \subfloat[\label{fig:perf_32_3_ecc_on}]{
\begin{centering}
    \includegraphics[width=12cm]{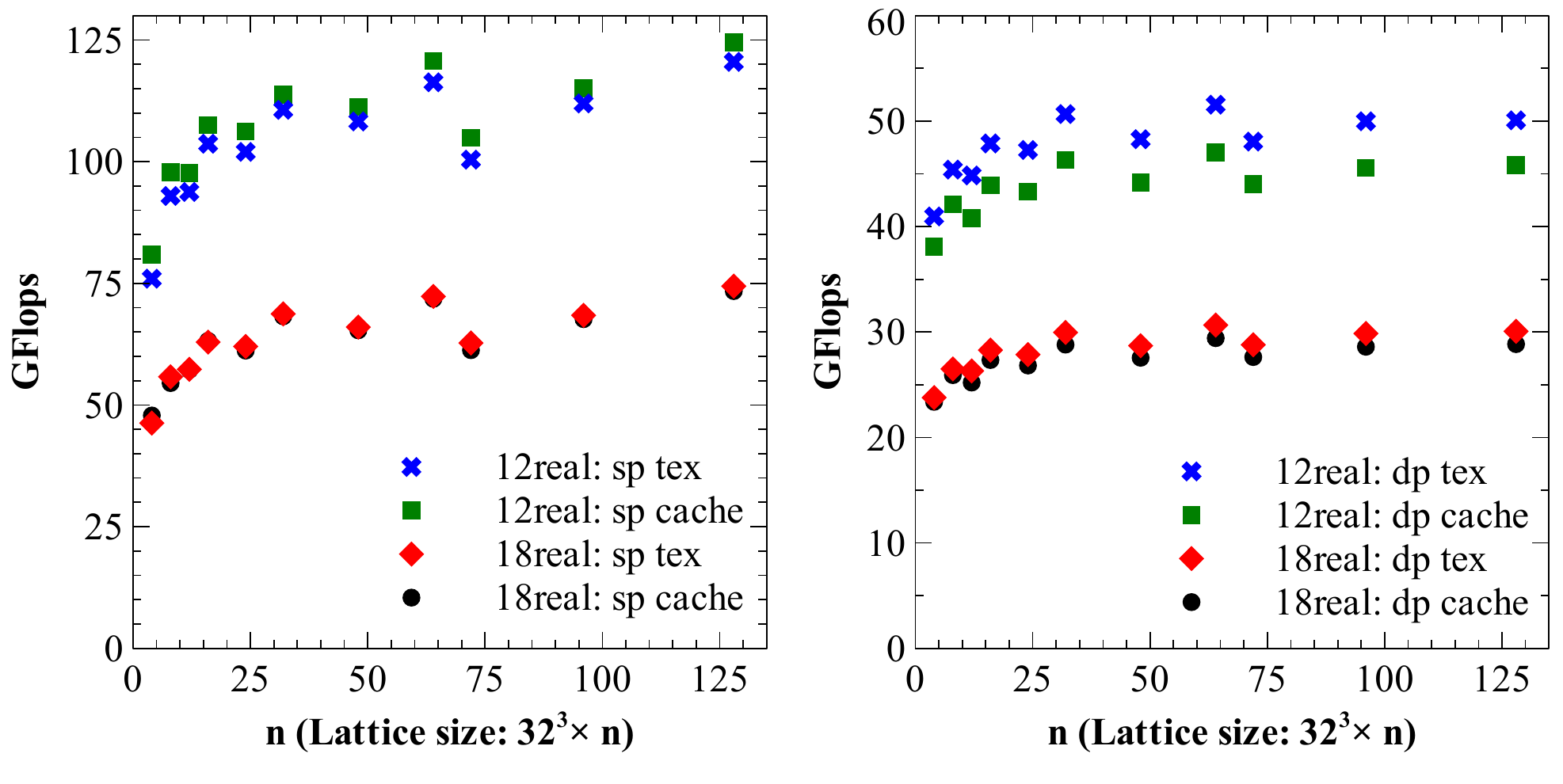}
\par\end{centering}}
\par\end{centering}
    \caption{Performance in GFlops, \cite{Cardoso:2012pv}. \protect\subref{fig:perf_32_3_ecc_off} with ECC Off and \protect\subref{fig:perf_32_3_ecc_on} with ECC On. sp: single precision, dp: double precision, 18real: full SU(3) matrix, 12real: 12 real parametrization, tex: using texture memory and cache: using L1 and L2 cache memory.}
    \label{fig:perf_32_3}
\end{figure}

In Fig. \ref{fig:perf_32_3}, we show the GPU performance, in GFlops, of the algorithm
using a 12 parameter reconstruction and the full (18 number) representation
in single and double precision. The GPU memory access is also compared, using L1 and L2 caches and the texture memory.
The best performance is achieved with ECC off, using texture memory and the 12 real number parameterization of the SU(3) matrix.

In order to compare the performance of the two codes, we use a $32^4$ lattice volume. The configurations have been generated using the standard Wilson gauge action, with three different values of $\beta$, 5.8, 6.0 and 6.2. In all runs, we set $\alpha =0.08$ and $\theta < 10^{-15}$.

\begin{figure}[!htb]
\begin{center}
\includegraphics[width=10cm]{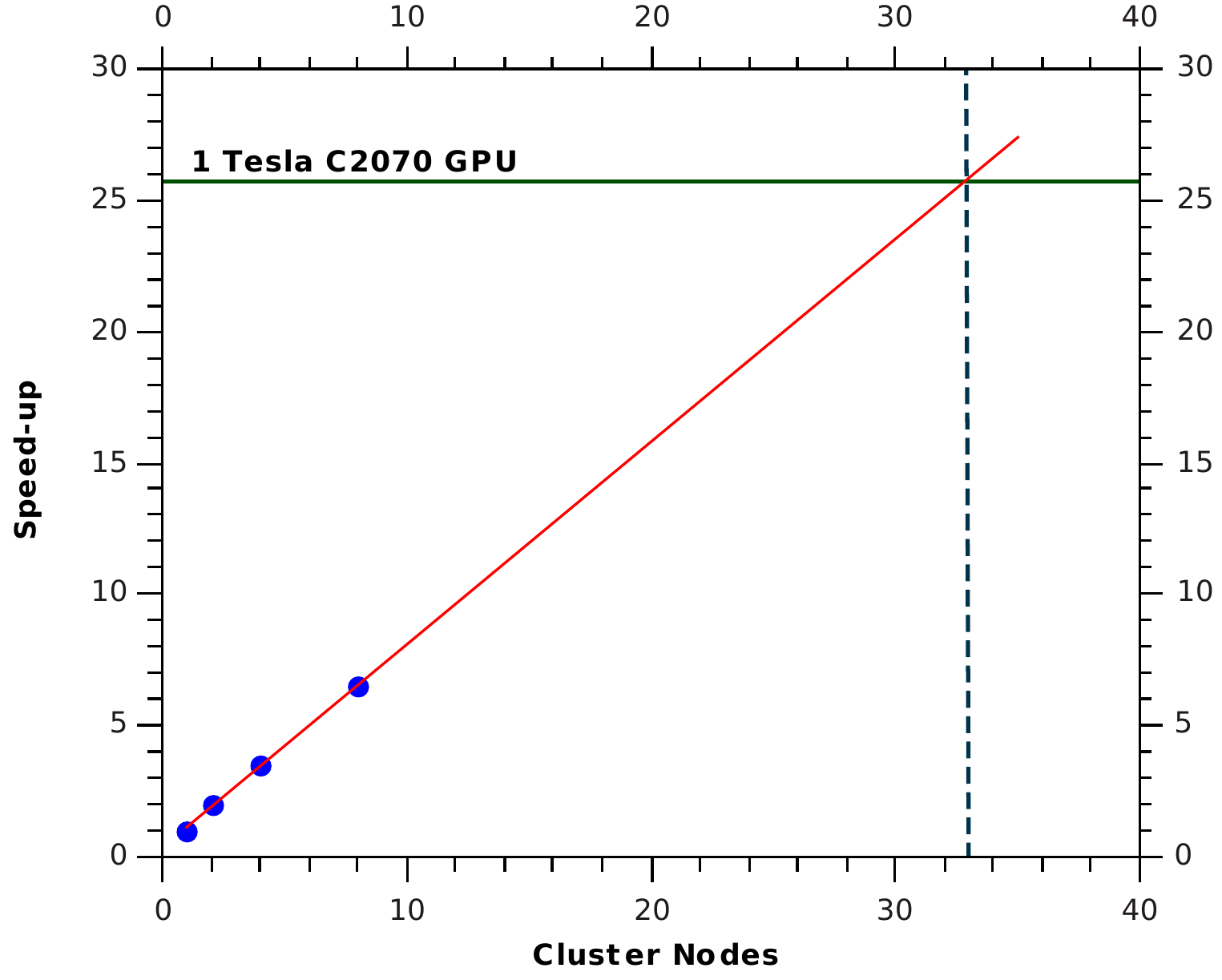}
\end{center}
\caption{Strong scaling CPU tested for a $32^4$ lattice volume and comparison with the GPU for the best performance, 12 real number parametrization, ECC Off and using texture memory in double precision, \cite{Cardoso:2012pv}. In Centaurus, a cluster node means 8 computing cores.}
\label{fig:cpuvsgpu}
\end{figure}

The CPU performance is compared with the best GPU performance, using 12 real parameterization, texture memory and ECC off, for a $32^4$ lattice volume in double precision.
The parallel CPU performance shows a good strong scaling behavior, Fig. \ref{fig:cpuvsgpu}, with a linear speed-up against the number of computing nodes. Nevertheless, the GPU code was much faster than the parallel CPU one: in order to reproduce the performance of the GPU code, one needs 256 CPU cores.

\section{Conclusion}

We have compared the performance of the GPU implementation of the Fourier accelerated steepest descent Landau gauge fixing algorithm using CUDA with a standard MPI implementation built on the Chroma library.
The run tests were done using $32^4$ $\beta = 5.8,6.0,6.2$ pure gauge configurations, generated using the standard Wilson action.

In order to optimize the GPU code, its performance was investigated on $32^3 \times n$ gauge configurations.
The runs on a C2070 Tesla show that, for a 4D lattice, the best performance was achieved with 2D plus 2D FFTs using \lstinline!cufftPlanMany()! and using a 12 real parameter reconstruction with texture memory. 

From all the runs using a C2070 Tesla GPU, peak performance was measured as 186/71 GFlops for single/double precision.
From the performance point of view, a run on a single GPU delivers the same performance as the CPU code when running on 32 nodes (256 cores), if one assumes a linear speed-up behavior, in double precision.

\section{Acknowledgments}

We thank B\'{a}lint Jo\'{o} and Michael Pippig for discussions about Chroma and PFFT libraries respectively. In particular, we thank Michael Pippig for extending his library for 4-dimensional FFTs. 

This work was partly funded by the FCT contracts  POCI/FP/81933/2007, 
CERN/FP/83582/2008, PTDC/FIS/100968/2008, CERN/FP/109327/2009, CERN/FP/116383/2010, CERN/FP/123612/2011, projects developed under initiative QREN financed
by UE/FEDER through Programme COMPETE.
Nuno Cardoso and Paulo Silva are supported by FCT under contracts SFRH/BD/44416/2008 and SFRH/BPD/40998/2007 respectively.
We would like to thank NVIDIA Corporation for the hardware donation used in this work via Academic Partnership 
program.

\bibliographystyle{elsarticle-num}
\bibliography{bib}

\end{document}